\documentclass[conference]{IEEEtran}
\usepackage{graphicx}
\usepackage{amsfonts}
\usepackage{amssymb,amsmath,amsthm}
\usepackage{euscript}
\usepackage{pstricks}
\usepackage{color}
\newtheorem{definition}{Definition}

\newtheorem{lemma}{Lemma}

\begin{document}
\title{Reachability Analysis of Time Basic Petri Nets:\\a Time Coverage Approach}
\author{\IEEEauthorblockN{Carlo Bellettini, Lorenzo Capra}
\IEEEauthorblockA{Department of Informatics and Communication\\
Universit\`a degli Studi di Milano, Italy\\
Email: \{bellettini, capra\}@dico.unimi.it}}
\maketitle

\begin{abstract}
We introduce a technique for reachability analysis
of Time-Basic (TB) Petri nets, a powerful formalism for real-time systems where time constraints are expressed
as intervals, representing possible transition firing times,
whose bounds are functions of marking's time description.
The technique consists of building a symbolic reachability graph relying on a sort of time coverage,
and overcomes the limitations of the only available analyzer for TB nets,
based in turn on a time-bounded inspection of a (possibly infinite) reachability-tree.
The graph construction algorithm has been automated by a tool-set,
briefly described in the paper
together with its main functionality and analysis capability.
A running example is used throughout the paper to sketch the symbolic graph construction.
A use case describing a small real system - that the running example is an excerpt from -
has been employed to benchmark the technique and the tool-set.
The main outcome of this test are also presented in the paper.
Ongoing work, in the perspective of integrating with a model-checking engine,
is shortly discussed.
\end{abstract}
\section{Introduction}
\label{sec:intro}
Time-Basic (TB) Petri nets \cite{UnifiedWay91} belong to the category of nets in which system time constraints
are expressed as numerical intervals associated to each transition, representing
possible firing instants, computed since transition's enabling time.
Tokens atomically produced by the firing of a transition are thereby associated to time-stamps
with values ranging over a determined set.
With respect to the well-known representative of this category, i.e.,
Time Petri nets \cite{Berthomieu91}, interval bounds in TB nets are linear functions
of timestamps in the enabling marking, rather than simply numerical constants.
TB nets thus represent a much more expressive formal model for real-time systems. 
The reachability analysis of TB nets is still recognized as an open problem (\cite{Hudak2010}).
Available analysis techniques and tools (e.g., \cite{Hudak2010,Cabernet93}) are based on inspecting
a finite portion of the potentially infinite reachability-tree generated by a TB net.
But for particular cases, only time-bounded properties can be inferred from
TB net's state-space exploration by using this kind of analyzers. 
The technique described in this paper tries to overcome this major limitation.
It relies on a symbolic reachability graph algorithm,
which is in turn based on a relative notion of time
and
a procedure verifying inclusion between symbolic states.
A particular state normalization, able to recognize and eliminate timestamp symbols
actually not influencing the model evolution,
permits in many cases building a sort of \emph{time coverage} finite graph.
The symbolic graph construction has been automated by a tool-set written
in Java.
The output is a structure enriched with information on edges which might be exploited
during property evaluation. The tool-set currently includes a module for
the automatic verification of properties expressed as conditions on markings.
As use case we'll use the gas burner example, that is widely used in literature as a representative of a small real system. A complete and formal description can be found in \cite{Ravn93}, and the corresponding  TB net model was introduced in \cite{IPTES-PDM41}. An excerpt will be used as running example to explain in a rather informal
way the essential points of symbolic graph construction. Only some relevant new core definitions  are formally given.

\section{TBnets}
\label{sec:tbnets}
Time Basic nets are Petri nets where each token is associated with a time-stamp representing the instant at which it has been created. In this paper we assume that the domain of timestamps is 
$\mathbb{R}^+$.
Each transition $t$ is associated with a \emph{time function} $f_t$ which maps a tuple $en$ of time-stamps, one for each
place in $^{\bullet}t$ (the pre-set of $t$), to a (possibly empty) set of $\mathbb{R}^+$ values. A marking $m$ maps
each place $p$ to a multi-set in $\mathbb{R}^+$. A tuple $en$ is said to be an enabling tuple for $t$ in $m$
if $m$ contains $en$ and $f_t(en) \neq \emptyset$. The set $f_t(en)$ represents the possible firing times of 
enabling tuple $en$. The firing of $(en,t)$ makes $en$ be withdrawn from $^{\bullet}t$, and a
new timestamp arbitrarily chosen among the values in $f_t(en)$ be created in all places in $t^{\bullet}$ (the post-set of $t$).

Hereafter a time function $f_t$ is defined by a pair of linear functions $[lb_t, ub_t]$, denoting interval bounds.
$lb_t, ub_t$ are in turn formally expressed in terms of
(a non empty set of) names of places in $^{\bullet}t$.
Time-functions are monotonic, i.e., the set of time-stamps associated with a tuple $en$ cannot contain a timestamp less than the maximum time-stamp associated with a token in $en$, denoted $enab$. We will keep such assumption implicit in the formal notation for time-functions.

\begin{figure*}[ht]
\begin{center}
{\includegraphics[width=0.6\textwidth]{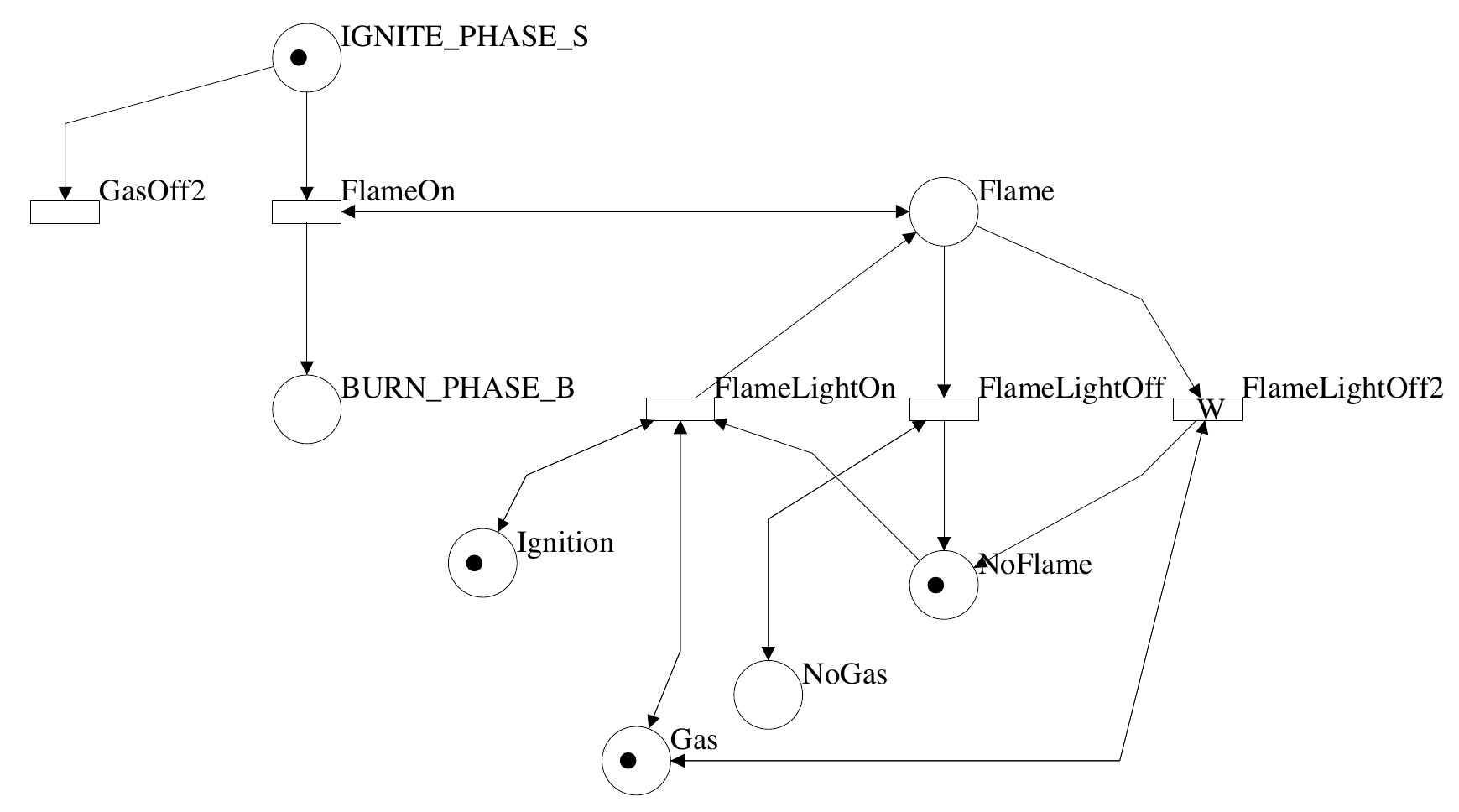}}
{\small
\vskip 0.5cm
{\scriptsize
\begin{tabular}{@{} ll @{}}
Initial marking & $IGNITE\_PHASE\_S\{T_0\}\,\,Ignition\{T_0\}\,\,Gas\{T_0\}\,\,NoFlame\{T_0\}$\\
Initial constraint & $0 \le T_0 \le 10$ \\
 & \\
\bf{FlameLightOff} & [$enab , NoGas + 0.1$] \\
\bf{FlameLightOff2} & [$enab, enab+100$] with weak time semantic \\
\bf{FlameLightOn} & [$enab+0.5 , enab+0.5$] \\
\bf{FlameOn} & [$IGNITE\_PHASE\_S + 0.01 , max(\{Flame+0.1 , IGNITE\_PHASE\_S + 0.01\})$] \\
\bf{GasOff2} & [$enab + 2 , enab + 2$]\\
\end{tabular}
}
}
\caption{running example}
\label{fig:example}
\end{center}
\end{figure*}

Consider the excerpt from the use case, depicted in Fig.~\ref{fig:example}.
It relates to the \emph{Ignite Phase}, just after the ignition transformer has been started and the gas valve has been opened. In this phase the controller must check if the flame has been lighted within a specific deadline, otherwise a recovery procedure that brings the system to \emph{Idle} has to be activated.
The flame turns on if there are \emph{Ignition} and \emph{Gas} (transition \emph{FlameLigthOn}), but it can turn off if no gas is supplied (transition \emph{FlameLigthOff}) or due to a fault, e.g. some wind (transition \emph{FlameLigthOff2}).
The time function associated with transition $FlameOn$ (representing the system passing to $burn state$ after recognizing that the flame has turned on)
can be interpreted as follows: $FlameOn$ cannot fire before 0.01 time units elapse since the appearance of a token
in place $IGNITE\_PHASE\_S$ (the minimum permanence time in $ignite state$) and implicitly not before the timestamp in place $Flame$. 
The firing time cannot exceed the maximum between the timestamp of the token in place $IGNITE\_PHASE\_S$ plus 0.01 time units and the time-stamp of the token in place \emph{Flame} plus 0.1 (i.e., the system recognizes the presence of a flame within this 0.1 units).
Noticeably, this is an example of constraint that cannot be directly expressed using Time Petri Nets formalism (\cite{Berthomieu91}). 

The set of time-stamps associated with a tuple by $f_t$ can be interpreted in at least two different ways, leading to different time semantics for each transition. A first interpretation states that an enabled transition $t$ $can$ fire at a value included in the set of possible firing times. Transitions with such semantics are referred to as \emph{weak} transitions. A second interpretation states that an enabled transition $t$ $must$ fire at a value included in the set of possible firing times unless it is disabled by the firing of any conflicting transition at a time no greater than the greatest firing time of $t$. Transitions with such semantics are referred to as \emph{strong} transitions.
Concerning the net in Fig.~\ref{fig:example}, the only weak transition is \emph{FlameLightOff2}. This permits us to express the $possibility$ that an event occurs within a given time interval.
In the case of Time Petri Nets the only possible semantics is strong.

In order to meet an intuitive notion of time, TB net firing sequences are restricted to the set of firing sequences whose firing times are monotonically non decreasing with respect to the firing occurrences.
However, the time of a firing may be equal to the enabling time of the tuple that belongs to the firing. Intuitively this means that an effect (the firing) can occur with no delay after the cause (that  enables it) is fulfilled. Therefore, it is possible to have sequences of firings where the time does not change. In practice, it is useful to restrict the attention to a subclass of TB nets, such that there exist no infinitely long firing sequences which take a finite amount of time (non Zenonicity).
\section{Time coverage reachability analysis}
\label{sec:technique}
The analysis technique presented in this paper extends the capability of the existing analyzer for TB nets \cite{Merlot93},
which uniquely permits the verification of bounded invariance and response properties,
through the inspection of a time-bounded symbolic reachability tree generated from a TB net.

The new technique aims at building a finite graph instead of an infinite tree
for a wide category of TB nets. A combination of three complementary
ideas is exploited. First, symbolic states are compared to check subset relationships.
For that purpose, using a consolidated approach, timestamp symbols no more
occurring on the marking description are eliminated
from the linear constraint associated to a symbolic state,
independently of how it has been reached.
Identifying subset relations between generated symbolic states (markings plus constraints),
is necessary for recognizing cyclic paths, but it is not enough in many situations.
As time progresses, periodic occurrences of equivalent conditions may be unrecognizable
simply due to their different offsets with respect to system's time zero.
This observation leads us dealing with the second aspect.
In the very common case a TB model contains no reference to \textit{absolute times} (i.e., not as offset respect to enabling timestamps)
in transition time functions, it is possible to remove any references
to the ``absolute zero'' from symbolic states.
This permits a periodic equivalent behavior to be recognized.
The cost is a lossy information about state displacement along absolute time.
We'll discuss this aspects in section~\ref{sec:prop}. 
Let us only point out that this kind of information could be recovered, if necessary,
in a second step by retracing only the path(s)
leading to the state of interest, or (at least partially) by combining the information on edges.
The third key feature of the technique is the introduction of the \emph{time anonymous} (TA) concept. This relates to the fact that in a symbolic state there may exist tokens
whose timestamp values can be \emph{forgotten}, as not influencing the evolution of a model. 
Several heuristics have been implemented, based on a mix of structural and state-dependent patterns,
each characterizing one such situation. 
This enhances the ability of merging states, and permits facing situations where the presence of
dead tokens could reintroduce a sort of \emph{symbolic} absolute zero, nullifying the achievements
at the previous points. 
Again, the cost to pay is a minor loss of information, as discussed later.
There is some resemblance with the approach used in the construction of (topological) coverage graphs:
the missing information is the exact timestamp of tokens instead of their exact number.
TA recognition might be also exploited to introduce a topological notion of coverage
for TB nets (section~\ref{sec:conclusion}).  

\subsection{Graph construction}
\label{sec:graphconstr}
In order to understand the rationale behind the symbolic reachability graph construction technique for TB nets,
we shall use once again the running example in Fig.~\ref{fig:example}.
Let us only introduce a few basic notions used in the sequel, referring to \cite{Ghezzi94}
(where the symbolic reachability tree for TB nets is defined) for a full formalization.

Let $TS = \{T_i\}$, $i \geq 0$, be the set of time-stamp symbols.
A symbolic state $S$ is a pair $\langle M, C\rangle$, where $M$ (called marking) maps each place $p$ to a multi-set
on $TS$, and $C$ is a (satisfiable) linear constraint defined on a subset of $TS$ symbols appearing in $M$.
We are considering a \emph{normal form} of $S$: if $M$ contains $k$ different $TS$ symbols,
they are $\{T_0,\ldots,T_{k-1}\}$, with the (implicit) assumption  $\forall i: 0\ldots k-2$, $C \Rightarrow T_i \leq T_{i+1}$.
Unless otherwise specified, we shall refer to this form.

A mapping $en_s\,:\, {^{\bullet}t} \rightarrow TS$ is said a \emph{symbolic
instance} of $t$ (the notation $(en_s,t)$ will be sometimes used).
$en_s$ will be formally denoted by a tuple of symbols.
A \textit{symbolic evaluation} of a linear function $g_t$ appearing in the formal definition of a time
function, denoted $g_t(en_s)$,
is obtained by replacing each occurrence of $p \in {^{\bullet}t}$ in the formal expression of $g_t$
with the associated symbol $\tau = en_s(p)$.
  
According to a (monotonic) weak time semantics, a pair $(en_s, t)$ is said a symbolic enabling 
in $S$ if $M$ contains $en_s$ and $C'$: $C \wedge lb_t(en_s) \leq T_{k} \leq ub_t(en_s) \wedge
T_{k-1} \leq T_{k} $ is satisfiable.
In other words, there exists at least one substitution $en$ of numerical values for $en_s$ that makes $C$
satisfiable and the set $f_t(en)$ non empty.

The firing of symbolic enabling $(en_s, t)$ produces the new symbolic state $S' : \langle M', C'\rangle$,
where $M'$ is obtained from $M$ by removing $en_s$ and putting a new symbol $T_{k}$ in all places in $t^{\bullet}$.
The state $S'$ represents all the possible TB net ordinary markings reachable from any marking
represented by $S$ by means of any firing correponding to $(en_s, t)$.

\subsection{Time-coverage graph}
\begin{figure*}[ht]
\centering
\includegraphics[width=0.7\textwidth]{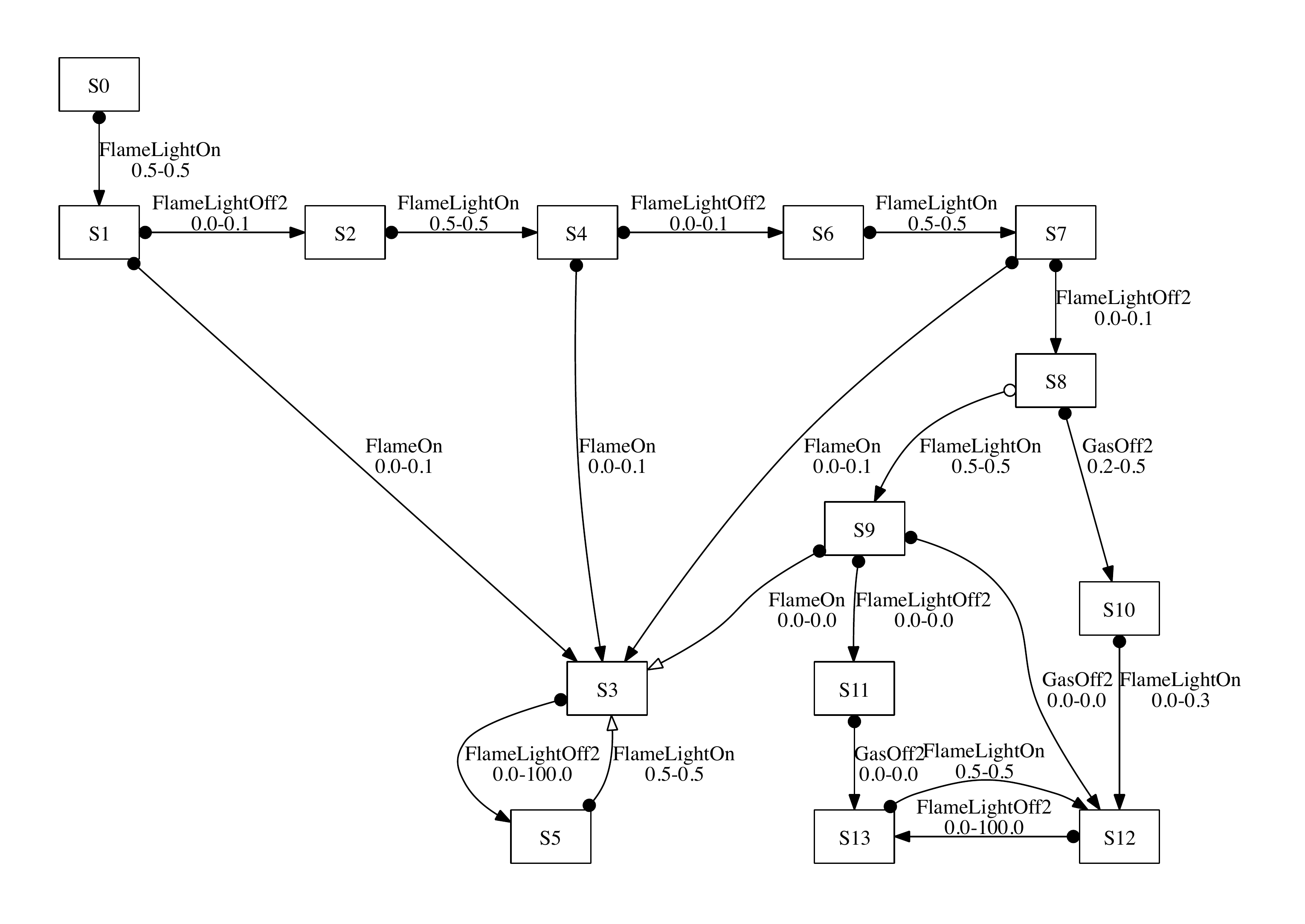}
\caption{sample reachability graph}
\label{fig:graph}
\end{figure*}

The time-coverage symbolic reachability graph generated by the running example, composed by 14 symbolic states,
is presented in Fig.~\ref{fig:graph}.\footnote{this picture has been automatically obtained by using GraphViz visualization software~\cite{graphviz} on the output generated from the tool-set.}.

The adopted notation for states is:
a square for symbolic states, a double square for symbolic states containing some deadlocks.
Concerning edges, the format of head and tail specifies the kind of relation between source and
target.

The \emph{normal} case, corresponding to a symbolic enabling,
is black head and tail, e.g., from $S0$ to $S1$:
considering any marking represented by $S0$ it is always possible to follow that edge and reach all the markings represented by $S1$.
 
Let us consider the symbolic state $S8$, formally described as follows:
$$ \left.\begin{array}{lll}M8 &: & Gas\{T_1\} \,\, IGNITE\_PHASE\_S \{T_0\}
\\ & & Ignition\{TA\} \,\, NoFlame\{TA\}
\\C8 &: & T_1 \ge T_0+1.5 \wedge T_1 \le T_0+1.8
\end{array}\right.$$
We can observe that, with respect to the original definition of symbolic state, a first extra time-stamp symbol
is present, \emph{TA} (time anonymous). This new symbol can occur only on the marking.
Postponing an intuitive explanation of when and how symbol \emph{TA} is
introduced in a symbolic state representation,
we can think of it as a token carrying on an unspecified time-stamp,
which has been shown unessential for the computation of transition firing times.
    
The ``candidates'' for symbolic enabling in $S8$ are: $(\langle T_0\rangle, GasOff2)$ and $(\langle TA,T_1,TA\rangle, FlameLightOn)$.
Firing times are computed by (symbolically) evaluating transition time functions, as explained above.
For \emph{GasOff2} the (only) inferred firing time is $\{T_0+2\}$.
Time function evaluation is slightly different for \emph{FlameLightOn}, due to the occurrence
of \emph{TA} in the pre-set tuple: this symbol is simply \textit{erased} during (symbolic)
evaluation: $enab = max(\{ TA,T_1,TA\}) \equiv max(\{T_1\}) = T_1$.
The inferred firing time in this case is $\{T_1+0.5\}$.

Since both transitions have a strong semantics, there are two additional constraints specifying that
the firing time of one cannot be greater than the (maximum) firing time of the other.\footnote{i.e., the set
of firing times of a (strong) transition in $S$ also depends on the enablings of the other strong transitions} They are $C_{\textrm{\tiny{GO2}}}:\, T_0+2 <= T_1+0.5$ and $C_{\textrm{\tiny{FLO}}}:\, T_1+0.5 <= T_0+2$, respectively.

Since both $C8 \wedge C_{\textrm{\tiny{GO2}}} \wedge T_2$ = $T_0+2$ and $C8 \wedge C_{\textrm{\tiny{FLO}}} \wedge T_2$ = $T_1+0.5$ are satisfiable, $(\langle T_0\rangle, GasOff2)$ and $(\langle TA,T_1,TA\rangle, FlameLightOn)$
are in fact symbolic enablings in $S8$.
However it is important to note that $C8 \Rightarrow C_{\textrm{\tiny{GO2}}} \wedge T_2 = T_0+2$, i.e.,  all the markings represented by $S8$ enable the transition \emph{GasOff2}. Instead $C8 \nRightarrow C_{\textrm{\tiny{FLO}}} \wedge T_2 = T_1+0.5$, i.e., only a subset of the markings expressed by $S8$ enable the transition  \emph{FlameLightOn}. This is highlighted in the graph by the white tail of the edge from $S8$ to $S9$.

Consider now the firing of $(\langle T_0\rangle, GasOff2)$: it only consumes tokens.
In such cases the symbolic firing rule slightly differs from the original one.
A second special symbol, $TL$ (Time Last), is introduced.
$TL$ can occur only on the constraint of a symbolic state and has an intuitive meaning:
it stands for the last firing time of the TB net and it permits
a correct interpretation of the model's time semantics.
\footnote{in this paper, when $TL$ is left implicit,
it is equal to the ``last'' generated timestamp $T_k$.} 
The reached symbolic state $S10$ is formally described as:

$$ \left.\begin{array}{lll}M10 &: & Gas\{T_1\} \,\,  Ignition\{T_A\} \,\, NoFlame\{T_A\} \\C10 &: & C8 \wedge T_2 = T_0+2 \wedge TL=T_2
\end{array}\right.$$
The normalization step eliminates symbols $T_2$ (the symbolic firing time) and $T_0$, as they occur only in $C10$, instead it leaves symbol $TL$. That results in (after a timestamp renaming):  
$$ \left.\begin{array}{lll}M10 &: & Gas\{T_0\} \,\,  Ignition\{T_A\} \,\, NoFlame\{T_A\} \\C10 &: & TL \geq  T_0+0.2 \wedge TL \leq T_0+0.5
\end{array}\right.$$
Another circumstance that causes the introduction of $TL$ symbol in a symbolic state representation is when
the maximum timestamp symbol $T_k$ is replaced with \emph{TA}. How identifying a Time Anonymous in a given symbolic state
is the next topic we treat. 

The graph in Fig.~\ref{fig:graph} contains two looping paths: between states $S3$ and $S5$, and between $S12$ and $S13$ respectively. That happens because in the extrapolated sub-model (Fig.~\ref{fig:example}), no expected actions are activated after the system exits the \emph{ignition phase} (e.g., closing the gas valve in the event of fail, or stopping ignition), so that an unbounded sequence of \emph{FlameLightOff2};\emph{FlameLightOn} is possible. 

The white head of the edge from $S5$ to $S3$ means that at least one of the ordinary
markings represented by $S3$ is not reachable by following that edge. 
This happens when a newly built symbolic state is recognized to be strictly included in an existing one.
What permits recognizing inclusion between states in this specific case
is the usage of \emph{Time Anonymous} timestamps. $S3$ is formally defined as:
$$ \left.\begin{array}{lll}M3 &: & Gas\{TA\} \,\, BURN\_PHASE\_B \{TA\}
\\ & & Ignition\{T_0\} \,\, Flame\{T_1\}
\\C3 &: &  T_1 \ge T_0 \wedge T_1 \le T_0+0.1
\end{array}\right.$$
Without using \emph{TA}s, its original definition ($S3'$) would be:
$$ \left.\begin{array}{lll}M3' &: & Gas\{T_0\} \,\, BURN\_PHASE\_B \{T_1\}
\\ & & Ignition\{T_0\} \,\, Flame\{T_1\}
\\C3' &: & T_1 \ge T_0 \wedge T_1 \le T_0+0.1
\end{array}\right.$$
Let us figure out what would be the model evolution from $S3'$, without introducing \emph{TA}.
After the firing sequence \emph{FlameLightOff2};\emph{FlameLightOn}\footnote{we omit in this description symbolic enablings, the TB net being safe} a state $S3''$ would be reached, defined in turn as:
$$ \left.\begin{array}{lll}M3'' &: & Gas\{T_1\} \,\, BURN\_PHASE\_B \{T_0\}
\\ & & Ignition\{T_1\} \,\, Flame\{T_1\}
\\C3'' &: & T_1 \ge T_0 +0.5 \wedge T_1 \le T_0+100.5
\end{array}\right.$$
Since $S3'' \not\subseteq S3'$ and $S3' \not\subseteq S3''$, there is no possibility to merge them and in fact the analysis tool would produce an infinite firing sequence.

Back to $S3$, we note it corresponds to $S3'$ but for holding \emph{TA} symbols in places $BURN\_PHASE\_B$ and $Gas$ instead of $T_1$ and
$T_0$, respectively.
Token $T_1$ in $BURN\_PHASE\_B$ however is not (and will never be) involved in any symbolic enabling because $BURN\_PHASE\_B$ has an empty postset, so it is immediately marked as \emph{TA}.
Token $T_0$ in $Gas$ instead is in the preset of transitions \emph{FlameLightOn} and \emph{FlameLightOff2}. As for \emph{FlameLightOn}, the tokens in place $Ignition$ and in place $Gas$ carry on the same timestamp, so either of them is enough to correctly evaluate transition's time function. As for \emph{FlameLightOff2}, the token in place $Gas$ carries on redundant information due to the simultaneous presence of $T_1$ in \emph{Flame}, that superseded it.

$S3''$ seems really different from $S3$, but nearly the same heuristics permits us to replace $T_0:\,BURN\_PHASE\_B$ ($T_i:\,p$ denotes the occurrence of a timestamp on a place) and $T_1:\,Gas$ with \emph{TA}s. That eliminates all the occurrences of $T_0$
from the marking. After timestamp renaming, we obtain the normal form:
$$ \left.\begin{array}{lll}M3'' &: & Gas\{TA\} \,\, BURN\_PHASE\_B \{TA\}
\\ & & Ignition\{T_0\} \,\, Flame\{T_0\}
\\C3'' &: & true
\end{array}\right.$$
However there is still a difference  with respect to $S3$:
places \emph{Ignition} and \emph{Flame} hold the same timestamp, but this boils
down to a condition already represented by $S3$ ($T_1 = T_0 \Rightarrow C3$), so $S3''$ is recognized as a state included in $S3$.

Notice that the other cycle on the graph, between $S12$ and $S13$,
is due to the adoption of a relative notion of time, i.e.,
it does not depend on the introduced \emph{TA} concept.    

An important setting of the legacy tool \cite{Cabernet93} was the \emph{time limit}, a positive interval time that guaranteed the finiteness of the symbolic reachability tree of a TB net. 
Upon elimination of absolute time references it has been substituted by a \emph{relative time limit}.
This positive interval specifies the maximum admissible distance between different timestamps in a state,
and allows one to deal with possibly infinite reachability graph.
The tool-set checks whether a symbolic state includes any ordinary states for which the distance between $TL$ and $T_0$ (the oldest meaningful timestamp) exceeds the time limit, marking that state as \emph{not to be expanded}.
The rationale behind is that reaching such a user defined limit might be a symptom of the presence of
unrecognized ``dead tokens'', reintroducing absolute time references.
If we analyzed the running example disabling \emph{TA} recognition,
the resulting graph would be infinite, unless a time limit is set.
Setting this limit to 3 (time units), 25 symbolic states would be generated:
13 already included in the presented graph,
the others corresponding to a partial unrolling of the loop between $S3$ and $S5$.

The output generated by the tool-set associates 
a couple of numerical values to edges of the graph,
corresponding to the minimum and maximum time distances from the source node to the target node.
This permits us to partially recover time relations between nodes that were lost due to
the removal of absolute times references from constraints.
In the following section we'll show how to exploit them.

\subsection{Formal Definitions}
Let us formalize some core concepts previously outlined, focusing in particular on \emph{TA} and coverage.

\begin{definition}[well-defined erasure]
\label{erasure}
Let $g_t$ be the formal expression of a linear function. The \textit{erasure}
of a set of symbols $E \subset {^{\bullet}t}$ from $g_t$ will be denoted ${g_t}_{[\neg E]}$.
${g_t}_{[\neg E]}$ is well-defined if it doesn't violate the arity of any operators occurring on $g_t$.
\end{definition} 
\noindent Consider for instance $t$, s.t. ${^{\bullet}t} = \{p_1,p_2\}$, and $f_t \,:$ $[max(\{p1,p2\}),p2+0.5]$,
where, $max\,:$ $2^{\mathbb{R}^+}\setminus \emptyset \rightarrow \mathbb{R}^+$, 
$+\,:\,\mathbb{R}^+,\mathbb{R}^+\rightarrow \mathbb{R}^+$. Then, the erasure ${f_t}_{[\neg \{p_1 \}]}$ is well-defined and
results in $[p2, p2+0.5]$, instead ${f_t}_{[\neg \{p_2 \}]}$ is not well-defined.

A symbolic instance of $t$ is a mapping $en_s\,:\, {^{\bullet}t} \rightarrow TS \cup \{\text{TA}\}$.
Let $en_s^{-1}(\tau)$ = $\{ p\}$, $en(p)=\tau$.

Let $\mathbf{R}(S)$ be the set of symbolic states reachable from $S$
\begin{definition}[valid \emph{TA}-replacement]
\label{def:TA-repl}
Given a state $S$, a timestamp occurrence $T_i : p$
is replaceable with $TA : p$ if and only if for each $S' = \langle M', C'\rangle\ \in \mathbf{R}(S)$
in which token $T_i : p$ is left (modulo timestamp renaming),
for each symbolic enabling $(en_s, t)$ in $S'$
s.t. $ en_s(p) = T_i$, ${f_t}_{[\neg \{p\}]}$ is a well-defined erasure and
\begin{center}
$C' \wedge max(\{TL,lb_t(en_s)\}) \leq ub_t(en_s) \Leftrightarrow C' \wedge max(\{TL,{lb_t}_{[\neg \{p\}]}(en_s)\}) \leq {ub_t}_{[\neg \{p\}]}(en_s)$
\end{center}
\end{definition}

The semantics of a symbolic state (possibly) including \emph{TA} is provided by the following coverage notion. 

\begin{definition}[symbolic state coverage]
\label{state-cov}
Let $S$ = $\langle M, C\rangle$ be a symbolic state.
An ordinary marking $m$ is covered by $S$ if and only if
it corresponds to a numerical substitution $\sigma$ of symbols occurring on $M$, s.t. $\sigma$ satisfies $C$ and
for each ordinary enabling $(en,t)$ in $m$,
for each symbolic tuple $(en_s,t)$ in $S$ s.t. $en$ is a numerical substitution of $en_s$,
\begin{itemize}
\item[a] ${lb_t}_{[\neg en_s^{-1}(\text{TA})]}$, ${ub_t}_{[\neg en_s^{-1}(\text{TA})]}$ are well defined
\item[b] ${lb_t}_{[\neg en_s^{-1}(\text{TA})]}(en) = {lb_t}(en) \wedge {ub_t}_{[\neg en_s^{-1}(\text{TA})]}(en) = {ub_t}(en)$
\end{itemize}
\end{definition}

The next lemma sets the relationship between ordinary and symbolic instances (state transitions).

\begin{lemma}
\label{lem:cover}
Let $m$ be covered by $S$. If $m [(en, t)> m'$, then there exists a symbolic enabling $en_s$,
s.t. $en$ is a numerical substitution of $en_s$, $S [(en_s , t) > S'$ and $m'$ is covered by $S'$
\end{lemma} 

Let us finally report as an example some heuristics, including the ones used by the algorithm to identify the \emph{TA} replacements 
 commented in the previous section. They identify, precisely speaking, a valid replacement of a timestamp occurrence
$T_k : p$ with $TA : p$, in $S$ = $\langle M, C\rangle$, according to definition \ref{def:TA-repl}.
\begin{enumerate}
\item $p^\bullet = \emptyset$
\item $\forall t \in p^\bullet \; \bigvee_i E_i$, where
\begin{itemize}
\item [$E_1$:]  $f_t$ does not refer to $p$, directly or by means of $enab$
\item [$E_2$:]  $f_t$ is in the form $[enab+c, enab+c']$ $\wedge$ $\exists p'\in{}^\bullet t$\\ $ M(p')=\emptyset$ $\; \vee \;$ $(\forall T_j \in M(p')\; C \Rightarrow T_j \geq T_k)$
\item [$E_3$:] \ldots
\end{itemize}   
\end{enumerate}

\section{Property Evaluation}
\label{sec:prop}

The symbolic (time coverage) reachability graph contains several exploitable information.

The tool recognizes deadlocks even if they are topologically hidden by the presence of outgoing edges. In fact if all the outgoing edges have a white tail, it is still possible that  a proper subset of the corresponding symbolic state is composed by deadlock marking. In the running example however no deadlock marking is reachable.

Disregarding time specification (i.e., considering only the number of tokens distributed over places),
the graph nodes exactly identify all the reachable (topological) markings: if a marking matches a symbolic node then there exists at least one path from the initial state to such a marking,
conversely if a marking matches no symbolic nodes, it is not reachable.
It is thereby possible to verify P-invariants from a specified marking.
In case of finite graph, it is possible to answer questions about maximum (minimum) number of tokens in some (combinations) of places.

In general, the set of ordinary (TB net) states represented by the sum of states of the symbolic graph built
from a TB net is a superset of the reachable ordinary states of the TB net.
In fact, the introduction of TA symbols causes a loss of information, because each TA covers a potential set
of timestamps. 
However, given a symbolic state $S = \langle M,C \rangle$ in which a set $\{T_i\}$ of time-stamp symbols occur on $M$, each numerical substitution of $\{T_i\}$ satisfying $C$ corresponds to the projection of  reachable ordinary states.
If we are interested in checking timing relations between token's timestamps on the states of the graph we
can get three different answers upon graph inspection: a positive one (e.g., there exists a node that satisfies the condition), a negative one (e.g., no nodes satisfy the condition), or a possibly positive.
For example, if we are looking for a state where a token in place \emph{Flame}
carries on a timestamp greater than the one in place \emph{IGNITION\_PHASE\_S},
state $S9$ provides us with a positive answer. Instead, if we are checking whether
places \emph{Gas} and \emph{Ignition} can ever hold the same timestamp the answer is may be
(the presence of TA in either places \emph{covers} that condition).

As for timing relations between token's timestamps in different markings, or between firing times in a transition
firing sequence, the symbolic graph permits identifying critical paths by combining the information on edges.
In particular, conservative bounds can be established. In the case they are not enough to exclude incorrect timing behaviors, it is possible to carry out a more accurate analysis by rebuilding a portion of the graph, retracing some critical paths and reintroducing absolute time references. For example, looking at the time information on edges, it is possible to establish that state $S10$ is not reachable from $S0$ in less than 1.7 time units.
We cannot directly infer that $S10$ is reachable in exactly 1.7 time units.
 
\begin{figure}[htbf]
\centering
\includegraphics[width=0.4\columnwidth]{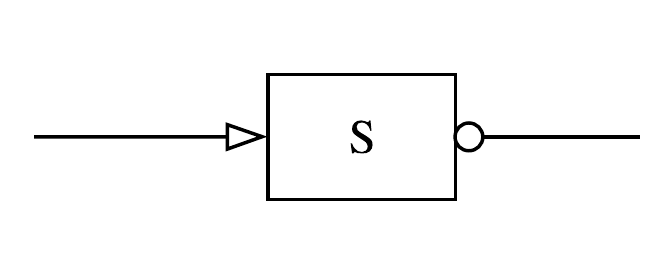}
\caption{Critical case for path feasibility}
\label{fig:EEpath}
\end{figure}

Concerning feasibility of firing sequences, the symbolic graph expresses all the possibilities (an ordinary firing sequence is matched by any firing sequence on the graph). A critical situation is presented in Fig.~\ref{fig:EEpath}. If we follow a white-arrow edge (meaning that we reach only a subset of the target state) and, from there, a white-tail edge (meaning that the transition is enabled only in a subset of the ordinary states represented by the node), there is still the possibility that this path actually is not feasible.
Also such critical paths could be retraced.
Let us stress (back to the reachability problem) that by construction, for every node on the graph there exists a path from the initial state to such a node formed exclusively by black-arrow edges.

The available tool's evaluation component is still very simple, its integration with some existing model checking engines is currently under investigation. However it already permits examining the input graph looking for interesting properties on topological definition of markings:
\begin{itemize}
\item existence of a state with a marking satisfying a constraint (i.e., a boolean combination of condition on the number of tokens in places)
\item maximum (minimum) value of an expression involving the number of tokens in places (possibly restricting the evaluation to markings satisfying a given constraint)
\end{itemize}

\section{Tool Architecture}
\label{sec:architecture}

\begin{figure}[htbf]
\centering
\includegraphics[width=\columnwidth]{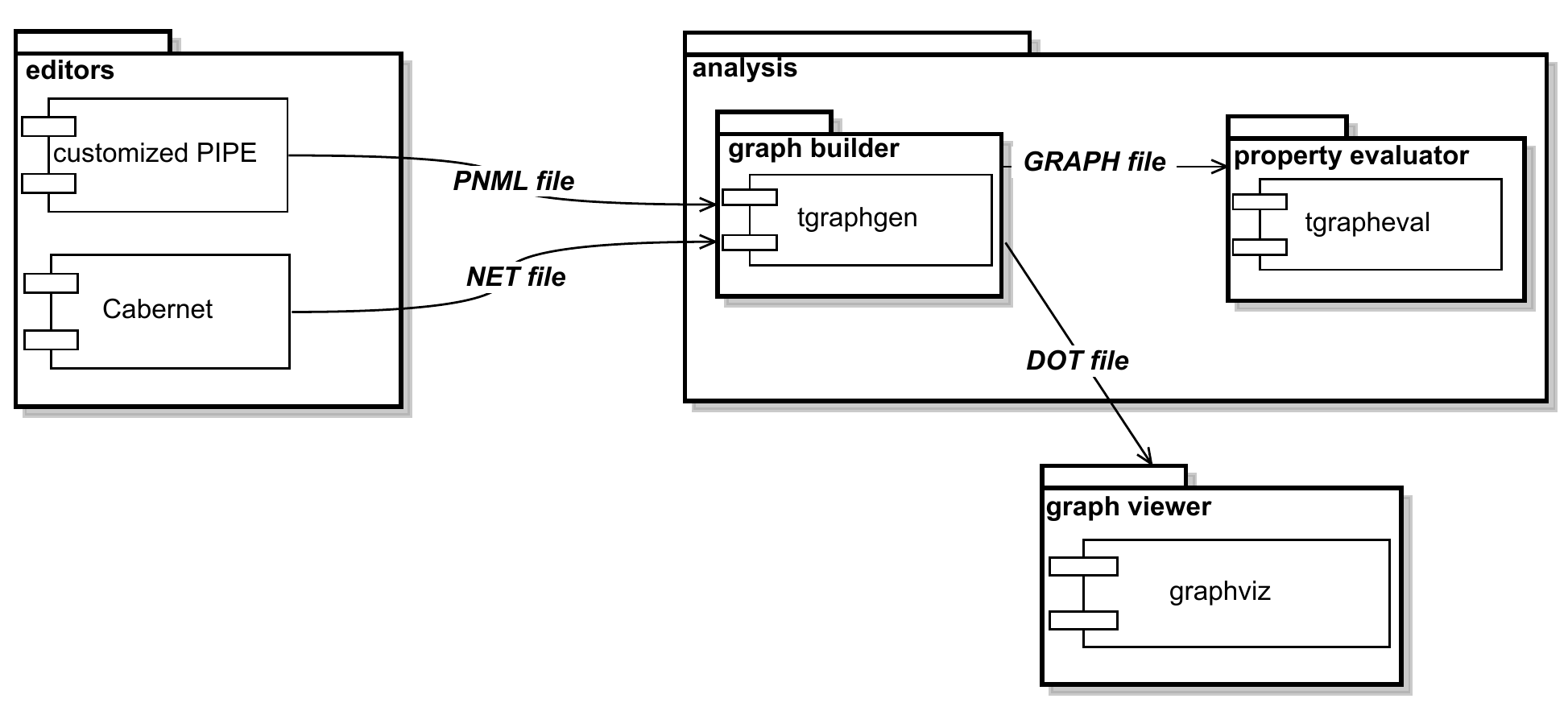}
\caption{reference architecture}
\label{fig:architecture}
\end{figure}

\begin{figure*}[htdb]
\centering
\includegraphics[width=0.8\textwidth]{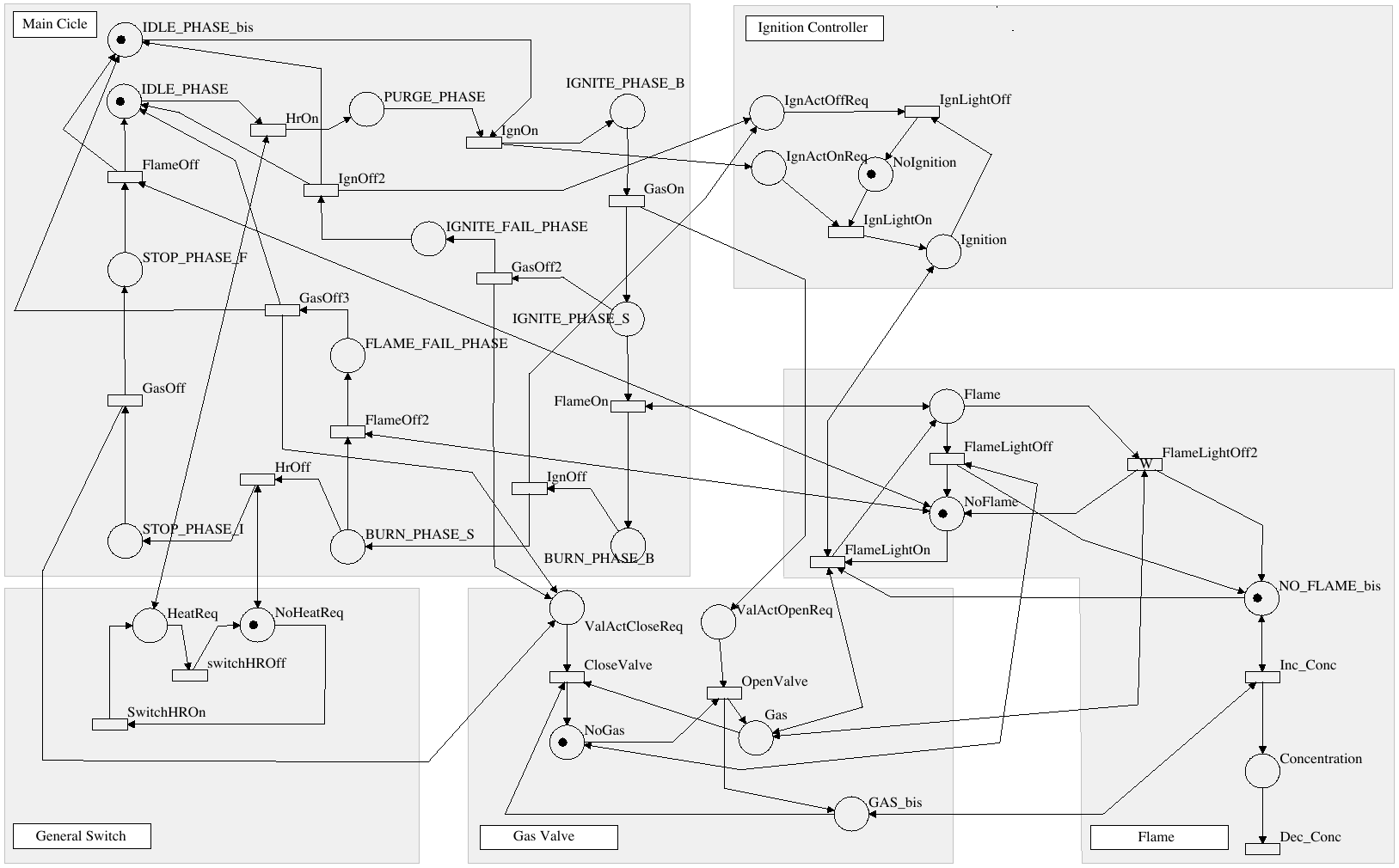}
{\footnotesize

{\bf Time-Functions:}
\[
\begin{array}{lll}
HrOn &  [ IDLE\_PHASE + 0.01, max (\{IDLE\_PHASE + 0.01 , HeatReq + 0.1\}) ] \\
IgnOn &  [ max(\{PURGE\_PHASE + 0.01,IDLE\_PHASE\_bis + 30\}),\\
& max(\{PURGE\_PHASE + 0.01,IDLE\_PHASE\_bis + 30\}) ] \\
GasOn &  [ enab + 0.01, enab + 0.1 ] \\
FlameOn &  [ IGNITE\_PHASE\_S + 0.01, max ( \{Flame+0.1 , IGNITE\_PHASE\_S+0.01\}) ] \\
IgnOff &  [ enab + 0.01, enab + 0.1 ] \\
HrOff &  [ BURN\_PHASE\_S + 0.01, max ( \{BURN\_PHASE\_S + 0.01 , NoHeatReq + 0.1 \}) ] \\
GasOff &  [ enab + 0.01, enab + 0.1 ] \\
FlameOff &  [ STOP\_PHASE\_F + 0.01, max ( \{STOP\_PHASE\_F+0.01, NoFlame+0.1 \}) ] \\
FlameOff2 &  [ BURN\_PHASE\_S + 0.01, max ( \{BURN\_PHASE\_S+0.01 , NoFlame+0.1 \}) ] \\
GasOff2 &  [ enab + 2, enab + 2 ] \\
IgnOff2 &  [ enab + 0.01, enab + 0.1 ] \\
GasOff3 &  [ enab + 0.01, enab + 0.1 ] \\
IgnLightOff &  [ IgnActOffReq + 0.2, IgnActOffReq + 0.2 ] \\
IgnLightOn &  [ IgnActOnReq + 0.2, IgnActOnReq + 0.2 ] \\
CloseValve &  [ ValActCloseReq + 0.2, ValActCloseReq + 0.2 ] \\
OpenValve &  [ ValActOpenReq + 0.2, ValActOpenReq + 0.2 ] \\
FlameLightOn &  [ max (\{Gas , Ignition \}) + 0.5, max (\{Gas , Ignition \}) + 0.5 ] \\
FlameLightOff &  [ enab, NoGas + 0.1 ] \\
SwitchHROn &  [ enab, enab + 10 ] \\
switchHROff &  [ enab + 120, enab + 120 ] \\
FlameLightOff2 &  [ enab, enab+100 ] \\
Inc\_Conc &  [ enab+0.1, enab+0.1 ] \\
Dec\_Conc &  [ enab+30, enab+30 ] \\
\end{array}
\]
}
\caption{Use case Net: Gas Burner}
\label{fig:gasburner}
\end{figure*}

The analysis technique described in this paper has been implemented as a command line tool written in Java. The tool architecture depicted in Fig.~\ref{fig:architecture} presents the various components that communicate by means of files.
The \emph{tgraphgen} module receives as input a Time Basic Petri net (either in the legacy file format used by the Cabernet tool, or in a PNML  format generated, for example, by a customized version of PIPE2 open source tool\cite{pipe2}). It generates as outputs the graph in binary format (used by the property verification module $tgrapheval$), and in an annotated DOT text format (used by the GraphViz tool).
The tool is going to be integrated as an analysis module in the customized PIPE open source tool. 
That will permit accessing all the functions by means of menu, and exploiting in an integrated environment consolidated structural analysis algorithms for the verification of the untimed part of TB nets (e.g., P/T nets invariant analysis).

\section{Use Case and Comparison with other tools}
\label{sec:comparison}
In order to make a comparison with the previous analysis techniques and tools available for TB nets, we consider now the complete gas burner example (Fig. \ref{fig:gasburner}), already analyzed in \cite{IPTES-PDM41}.

The main critical parameter of the system was identified in the maximum concentration value of uncombusted gas. With the old analyzers  it was only possible to do an approximate analysis, by verifying the safety requirement having fixed a time threshold \cite{IPTES-PDM41}, or by building a small part of the reachability tree able to invalidate the property \cite{Calzolari}. 
A significant improvement of the new tool-set against the old analyzer is that it is now possible to compute the upper bound for such a parameter.

Table~\ref{tab:results} reports the outcomes of the analysis on the use case. In particular the considered parameter has been measured with three versions of the net. They differ in the time granularity used for the uncombusted gas process, i.e., the time function of the transition $Inc\_Conc$. The first thing to note is however that the analysis result is coherent in the various situations, identifying the maximum amount of uncombusted gas as corresponding to a leaking period of two seconds.

The test has been performed on a Toshiba Notebook with 2.4Ghz Intel Core 2 Duo processor and 4GB of memory. The operating system is Ubuntu 10.10 and the Java Virtual Machine is OpenJDK IcedTea6 1.9.5.

On the table we report also the number of states of the final reduced graph against the overall number of states
generated by the algorithm, and the execution times.  
Even if the sample is too little, a first interpolation suggests that a likely trend is quadratic
(with a small constant factor). 

\begin{figure}[htbf]
\centering
\includegraphics[width=\columnwidth]{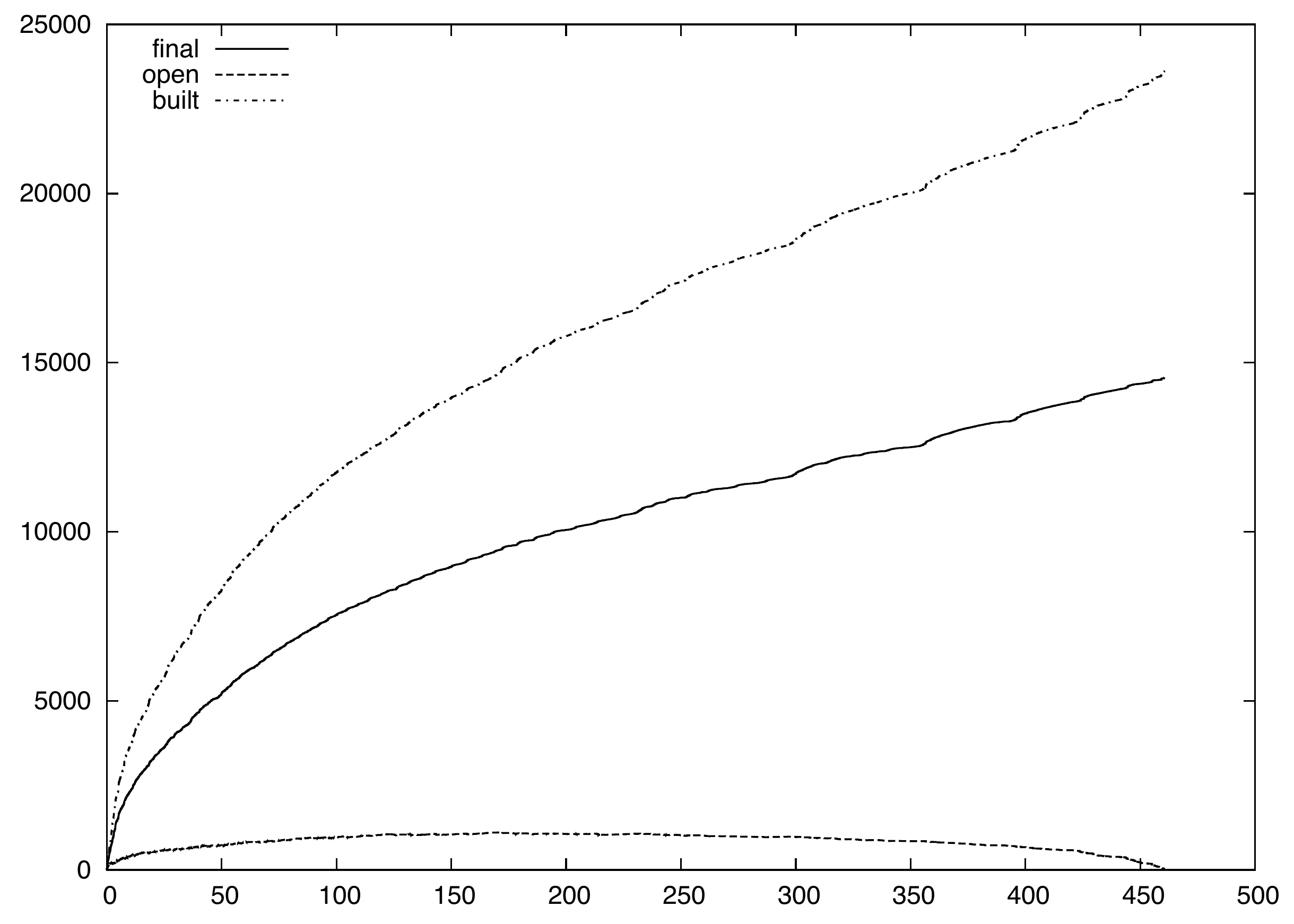}
\caption{state creation advancement}
\label{fig:01graph}
\end{figure}

In Fig.~\ref{fig:01graph} some profiling data -relating the 0.1 time granularity version of the model-
are presented.
On the x axis there is the execution time expressed in minutes, on the y axis there are the number of built nodes, of
reduced (final) nodes, and of nodes ready to be processed, respectively.
This picture is important for two reasons: first it shows that the performance degradation of state construction process is very small (the number of states created is pretty much constant in time after an initial burst); second, it supports the idea that a parallel (distributed) version of the graph builder, currently under development, should substantially improve the performances (the front of expansion remaining consistently wide).

\begin{table}[htdp]
\caption{Use Case Analysis results}
\begin{center}
\begin{tabular}{@{} |c|c|c|c|c| @{}}
\hline \hline
$Inc\_Conc$ granularity & max(Conc) & \#  [final/built] states & exec. time\\
\hline \hline
0.5 &  4 & 865/1217  & $\approx 75 secs$ \\
\hline
0.25 &  8 & 2233/2983 & $\approx 400 secs$ \\
\hline
0.1 &  20 &  14563/23635  & $\approx 7.5 hrs$ \\
\hline
\end{tabular}
\end{center}
\label{tab:results}
\end{table}

\section{Conclusion and future works}
\label{sec:conclusion}

The analysis technique presented in this paper overtakes the existing available
analysis technique for Time Basic Nets (a very expressive timed version of Petri nets) because it permits the building of a sort of (symbolic) time-coverage reachability graph keeping interesting timing properties of the nets.
In particular the introduction of the concept of \emph{time anonymous} timestamps, allows for a
major factorization of symbolic states.
An extension of the technique that further exploits the time anonymous concept in order to deal with topologically unbounded nets (by means of a coverage of \emph{TA} tokens, i.e., a sort of $\omega_{\textrm{\tiny{TA}}}$) is under definition.

\bibliographystyle{IEEEtran}
\bibliography{PN2011}

\end{document}